# Materialized View Replacement using Markov's Analysis

Partha Ghosh
A.K. Choudhury School of Information Technology
University of Calcutta
Kolkata, India
pghosh44@gmail.com

Soumya Sen
A.K. Choudhury School of Information Technology
University of Calcutta
Kolkata, India
iamsoumyasen@gmail.com

*Abstract*—Materialized view is used in large data centric applications to expedite query processing. The efficiency of materialized view depends on degree of result found against the queries over the existing materialized views. Materialized views are constructed following different methodologies. Thus the efficacy of the materialized views depends on the methodology based on which these are formed. Construction of materialized views are often time consuming and moreover after a certain time the performance of the materialized views degrade when the nature of queries change. In this situation either new materialized views could be constructed from scratch or the existing views could be upgraded. Fresh construction of materialized views has higher time complexity hence the modification of the existing views is a better solution. Modification process of materialized view is classified under materialized view maintenance scheme. Materialized view maintenance is a continuous process and the system could be tuned to ensure a constant rate of performance. If a materialized view construction process is not supported by materialized view maintenance scheme that system would suffer from performance degradation. In this paper a new materialized view maintenance scheme is proposed using markov's analysis to ensure consistent performance. Markov's analysis is chosen here to predict steady state probability over initial probability.

## I. INTRODUCTION

A database system may maintain materialized views for faster query processing. These materialized views are located either in primary memory (cache memory or main memory) or in secondary memory. Such a system works fine when it could answer majority of the user queries from the materialized views resided in primary memory. As over the time new queries come to the system materialized views in primary memory may not answer all of them. Thus over the time the performance of the system deteriorates. This research work is focused to find the materialized view from secondary memory which could answers query efficiently in this context.

In these types of systems generally a set of materialized views are formed using certain algorithms. This is a time consuming process. Hence any system which employs materialized view must be supported by the materialized view maintenance scheme. One of the solutions could be running the materialized view creation algorithm after a certain time instant. However as mentioned earlier, materialized view creation is a time consuming process hence it should be avoided by some other schemes. A scheme which could replace the materialized view from primary memory which is not being used for a long time, with a new view from secondary memory which is likely to be used frequently by the queries is matter of interest in this context. In this research work researchers try to identify a method which incorporates this.

In order to do this, researchers assume that there are several views residing in the primary memory as well as in the secondary memory .The goal of this project is to find an important view that is currently resides in the secondary memory and needs to bring in primary memory.

The views which are resided in secondary memory, for each of them at first "Initial Probability" is calculated to identify the importance of each view. But the nature of the query is uncertain, so it may possible that after a long transmission period this initial probability changes. Hence a stable probability calculation is required to transfer views from secondary to primary memory for answering the future queries directly from the materialized views resided in the primary memory. The second part of this paper computes "Steady State Probability Calculation" using Markov's analysis. The steady state probabilities are average probabilities that the system will be in a certain state after some transition period.

## II. RELATED WORK

The success of business sensitive processes depends on the capability of taking decisions quickly and incorporating them fast and dynamically. Building materialized view in this context help in quick decision making. Materialized View is created and cached in fast memory to expedite the query processing and query optimization [1]. It is applicable in large data centric applications like Database, Data Warehousing, Data Mining etc. Many researches have been done in this field to create materialized view based on heuristic algorithm [4], genetic algorithm [5] etc. A numeric scale is proposed in [2], [3] to show the association among participating attributes in the executed query set by analyzing the association of different attributes. [2] does it using standard deviation and [3] performs using linear regression. [3] also shows how to create materialized view from the numeric scale. These works however doesn't mention how these views will be maintained in future. So over the time as new queries come to the system, materialized views of the primary memory may not answer all of them. However, the main concern of creating materialized views is to ensure availability of higher amount of user requested data directly from materialized views. Thus the performance of the system deteriorates over the time. So after view creation next thing is to address is maintenance of views. In [10] a logical maintenance technique is being proposed which tries to compensate the effect of deferred maintenance. A system is proposed in [10] to distinguish between the maintenance of logical contents and physical structure. It results in support for concurrent high update rates and immediate, index-based query processing with correct transaction semantics. Evolvable View Environment (EVE) (EVE) [9] was proposed for view refreshment. In EVE [9], the view synchronizer rewrites the view definitions by replacing view components with suitable components from other information systems. It proposes strategies to address this incremental adaptation of the view extent after view synchronization. Incremental maintenance technique proposed in [6], which is illustrated in terms of algorithm and the experimental result shows the cost reduction in view maintenance. A new data model named as Chronicle data model [7] permits the capture, within the data model, of many computations common to transactional data recording systems. The central issue of this model is incremental maintenance of materialized views is time independent of the size of the recorded stream. This Chronicle data model [7] measure the complexity of a chronicle model by the complexity of incrementally maintaining its persistent views, and develop languages that ensure a low maintenance complexity independent of the sequence sizes. A lazy view maintenance scheme for social networking known as CAMEL [8] CAMEL employs the existing view maintenance techniques of incremental maintenance, lazy maintenance, and control table. Additionally CAMEL optimizes view maintenance performance by pushing the top-k operation down to before join operations and by constructing a reverse index. The experimental results show that CAMEL is faster than the method of eager view maintenance.

However none of these methods deal with the probabilistic nature of the incoming queries. The incoming queries change over the time. Hence the nature of the query is uncertain. In order to address this type of situation a steady state probability calculation is desirable. This research work is focused to identify a steady state probability that gives a better prediction of future state of the system.

## III. METHODOLOGY PROPOSED

In this methodology the proposed algorithm runs separately on secondary memory and also on primary memory. In secondary memory the algorithm finds out which view is important and in primary memory it finds out which one is less important. At first this runs on secondary memory and thereafter on primary memory. In case of secondary memory the view for which steady state probability value is best is selected to bring into primary memory and in case of primary memory the view for which steady state probability value is lowest is considered for replacement.

The proposed methodology has two parts. The First part calculates the "Initial Probability". However after a long transmission period this initial probability may be changed, thus this initial probability is uncertain. Hence a stable calculation of probability is required to transfer views from secondary to primary memory. The second part of this paper computes "Steady State Probability Calculation" using Markov's analysis.

### A. Selection of Best View from Secondary Memory

**Input:** This process starts by accepting a set of materialized views reside in secondary memory and a numbers of query set which fetch these views in case of a miss in primary memory.

**Step 1**: **Initial Probability Calculation**

Nature of query is uncertain, that is it may possible that the incoming queries are currently hitting into a view that resides in secondary memory, but the new queries may not hit that particular view of secondary memory, instead they are fetching some other views in secondary memory.

So based on the input, at every instance, if queries are presently hitting in to the $i^{th}$ view, hits into the $j^{th}$ view a (m × n) *View Hit Matrix* (VHM) is formed, where m and n denotes the number of distinct queries and number of views respectively. That is, if queries are using an initial view $V_i$ and continue hitting it for the next time then the corresponding cell value of $V_i$ will be "HIT", the moment it misses $V_i$ and hits any other view $V_j$ the iteration stops and calculate the probability of "If queries are initially hitting in $V_i$, the chance that the next query will hit in $V_j$ ". This process continues for each view resides in the secondary memory.

**Step 2: Steady State Probability Calculation:**

This process starts by accepting the probability of if queries hit in a specific view for the first time then what is the probability that the next query will hit in that particular view and also the probability that the next query will hit the other views. It is represented by a (n × n) matrix called "Initial Probability Matrix". Then the $n^{th}$ future state of the system is calculated by the Markov's analysis. An one dimensional unit matrix called a transition matrix 'T' is introduced here. This calculates the future states of the system by multiplying present state with Initial Probability Matrix.

A symbol $VN_{vn}$ (i) is introduced here. Where VN= probability of hit at present; vn= initial starting state; i = $i^{th}$ future period.

Now mathematically we can say that the probability of a query hitting in the $1^{st}$ view for the first time, given that the query hits in the $1^{st}$ view is 1.

This probability is represented as matrix in following form $[V1_{v1}(1) \quad V2_{v1}(1) \quad V3_{v1}(1)] = [1.0 \quad 0.0 \quad 0.0]$, ( 3 views are assumed in the secondary memory).

Now at this stage following the Markov's analysis, next system state probability is calculated by multiplying transition matrix 'T' with "Initial Probability Matrix". The resultant matrix is again multiplied with "Initial Probability Matrix" to get the probability of future. It is repeated until a steady state probability is achieved.

After getting steady state probability, we transfer the view that has the highest probability from secondary device to primary device.

**Algorithm:**

**Start**

**Step 1:** /* This method accepts the view that is being hit by recent input query. It returns the probability of the next query that will hit into the same view and also returns the name of the view and corresponding probability, which is used by the current query in case not answered by the present view*/

Call method Initial_Probability_Calculation ($i^{th}$_View)

**Step 2:** /* The initial probability is being calculated and stored in Initial Probability Matrix */

Call method Initial_Probability_Matrix( )

**Step 3:** /* This method is being used for steady state probability calculation. It multiply the present state with Initial Probability Matrix repeatedly until a steady state probability comes /*

Call method Steady_State(Previous_Transition_Matrix, Probability_Matrix)

**End**

**Algorithm Initial_Probability_Calculation ($i^{th}$_View)**

/* Cal_Hit( ) is a method that finds which view is been hit by the present query */

**Start:**

```
While (True)
        Present_hit= Cal_Hit( )
        If Present_hit = i^th view   Then
        Total= Total + 1
        Else
                Break
        End If
End While
I_th_Probability = Total / (Total + 1)
J_th_ Probability = 1 / (Total + 1)
Return(J_th_View, I_th_Probability, J_th_Probability)
```

**End**

**Algorithm Initial_Probability_Matrix( )**

**Start:**

```
For  i=1 to n
  For  j= 1 to n
   Probability_Matrix[i][j] = probability of hitting j^th view if
                               previously hits  in i^th view
  End For
End For
End
```

**Algorithm Steady_State(Previous_Transition_Matrix, Probability_Matrix)**

**Start:**

Present_Transition_Matrix = Previous_Transition_Matrix × Probability_Matrix

If Present_Transition_Matrix = Previous_Transition_Matrix Then

    Return

Else

    Steady_State(Present_Transition_Matrix ,Probability_Matrix )

End If
**END**

*B. Selection of Worst View from Primary Memory*

**Input:** This process starts by accepting a set of materialized views resided in primary memory and a numbers of query set.

The same steps are to be repeated like subsection A (previous section). This step is required in case when primary memory is full or not has enough space to cater a new view from secondary memory.

After execution of the algorithm the view with lowest value in steady state probability is selected for replacement.

## IV. AN ILLUSTRATIVE EXAMPLE

In this section an example is shown on the views located in secondary memory. The execution of the proposed algorithm chooses the most important views from secondary memory. We assume that there are 3 views in the secondary memory. Let starts from $1^{st}$ view $V_1$, that is the query presently hits in the view V1. The snap shots are as follows.

|       | $V_1$ | $V_2$ | $V_3$ |
|-------|-------|-------|-------|
| $Q_1$ | HIT   | MISS  | MISS  |
| $Q_2$ | HIT   | MISS  | MISS  |
| $Q_3$ | HIT   | MISS  | MISS  |
| $Q_4$ | MISS  | HIT   | MISS  |

And,

|       | $V_1$ | $V_2$ | $V_3$ |
|-------|-------|-------|-------|
| $Q_1$ | HIT   | MISS  | MISS  |
| $Q_2$ | HIT   | MISS  | MISS  |
| $Q_3$ | MISS  | MISS  | HIT   |

So, the probability that if initially queries are hitting in the 1st view then it will hit in the 1st view or the 2nd view for the next time is 3/4 and 1/4 respectively.

Again, the probability that if initially queries are hitting in the 1st view then it will hit in the 1st view or in the 3rd view in the future is 2/3 and 1/3 respectively.

So, from this two snapshot we can find the probability that if initially queries are hitting in the 1st view then it will hit in the 1st view or in the 2nd view or in the 3rd view in the future is,

( [3/4 + 2/3, 1/4 + 0, 0 + 1/3] ) / 2

= [ 17/24 , 1/8 , 1/6 ]

Similarly, for $V_2$ and $V_3$, let these are [ 1/5 , 7/10 , 1/10 ] and [ 1/10 , 1/10 , 4/5 ] respectively.

So, the "Initial Probability Matrix" is:

|       | $V_1$ | $V_2$ | $V_3$ |
|-------|-------|-------|-------|
| $V_1$ | 17/24 | 1/8   | 1/6   |
| $V_2$ | 1/5   | 7/10  | 1/10  |
| $V_3$ | 1/10  | 1/10  | 4/5   |

Now the probability of a query hitting in the 1st view at present, given that the query hits in the 1st view is 1.0. These probabilities can be arrange in matrix form as follows.

[$V1_{v1}(1)$  $V2_{v1}(1)$  $V3_{v1}(1)$] = [1.0  0.0  0.0]

This matrix is multiplied with "Initial Probability Matrix" to get the probability of next time.

i.e., [$V1_{v1}(2)$  $V2_{v1}(2)$  $V3_{v1}(2)$]

= [1.0  0.0  0.0] ×

|       | $V_1$ | $V_2$ | $V_3$ |
|-------|-------|-------|-------|
| $V_1$ | 17/24 | 1/8   | 1/6   |
| $V_2$ | 1/5   | 7/10  | 1/10  |
| $V_3$ | 1/10  | 1/10  | 4/5   |

= [ 0.708 , 0.125 , 0.167 ]

The resultant matrix [0.708 , 0.125 , 0.167] is again multiplied with "Initial Probability Matrix" to get the probability of next time. We repeat this until a steady state probability comes.

i.e., [$V1_{v1}(3)$  $V2_{v1}(3)$  $V3_{v1}(3)$]

= [0.708, 0.125, 0.167] ×

|       | $V_1$ | $V_2$ | $V_3$ |
|-------|-------|-------|-------|
| $V_1$ | 17/24 | 1/8   | 1/6   |
| $V_2$ | 1/5   | 7/10  | 1/10  |
| $V_3$ | 1/10  | 1/10  | 4/5   |

= [0.543 , 0.193 , 0.264]

Next, [$V1_{v1}(4)$  $V2_{v1}(4)$  $V3_{v1}(4)$]

= [0.543, 0.193, 0.264] ×

|       | $V_1$ | $V_2$ | $V_3$ |
|-------|-------|-------|-------|
| $V_1$ | 17/24 | 1/8   | 1/6   |
| $V_2$ | 1/5   | 7/10  | 1/10  |
| $V_3$ | 1/10  | 1/10  | 4/5   |

= [ 0.449, 0.229, 0.322]

Next, [$V1_{v1}(5)$  $V2_{v1}(5)$  $V3_{v1}(5)$]

= [0.449, 0.229, 0.322] ×

|       | $V_1$ | $V_2$ | $V_3$ |
|-------|-------|-------|-------|
| $V_1$ | 17/24 | 1/8   | 1/6   |
| $V_2$ | 1/5   | 7/10  | 1/10  |
| $V_3$ | 1/10  | 1/10  | 4/5   |

= [ 0.396, 0.247, 0.357]

Next, [$V1_{v1}(6)$  $V2_{v1}(6)$  $V3_{v1}(6)$]

= [0.396, 0.247, 0.357] ×

|       | $V_1$ | $V_2$ | $V_3$ |
|-------|-------|-------|-------|
| $V_1$ | 17/24 | 1/8   | 1/6   |
| $V_2$ | 1/5   | 7/10  | 1/10  |
| $V_3$ | 1/10  | 1/10  | 4/5   |

= [ 0.365, 0.258, 0.377]

Next, [$V1_{v1}(7)$  $V2_{v1}(7)$  $V3_{v1}(7)$]

=[0.365,0.258,0.377]×

|       | $V_1$ | $V_2$ | $V_3$ |
|-------|-------|-------|-------|
| $V_1$ | 17/24 | 1/8   | 1/6   |
| $V_2$ | 1/5   | 7/10  | 1/10  |
| $V_3$ | 1/10  | 1/10  | 4/5   |

=[ 0.347, 0.264, 0.389]

Next, [$V1_{v1}(8)$  $V2_{v1}(8)$  $V3_{v1}(8)$]

=[0.347,0.264,0.389]×

|       | $V_1$ | $V_2$ | $V_3$ |
|-------|-------|-------|-------|
| $V_1$ | 17/24 | 1/8   | 1/6   |
| $V_2$ | 1/5   | 7/10  | 1/10  |
| $V_3$ | 1/10  | 1/10  | 4/5   |

=[0.337, 0.267, 0.396]

Next, [$V1_{v1}(9)$  $V2_{v1}(9)$  $V3_{v1}(9)$]

=[0.337,0.267,0.396]×

|       | $V_1$ | $V_2$ | $V_3$ |
|-------|-------|-------|-------|
| $V_1$ | 17/24 | 1/8   | 1/6   |
| $V_2$ | 1/5   | 7/10  | 1/10  |
| $V_3$ | 1/10  | 1/10  | 4/5   |

=[0.331, 0.269, 0.400]

As the next state is almost the same so the steady state probability is approximately [0.33, 0.27, 0.40] (correct up to two decimal places). So the most important view is $V_3$ that needs to be in the main memory.

## V. CONCLUTION AND FUTURE WORK

This paper proposes a novel methodology to maintain Materialized View using Markov's analysis. This method is flexible, dynamic and independent of application areas. Moreover this methodology does not have any implicit or explicit assumptions. Hence this methodology is independent of the size or type of database and the corresponding areas of applications. Thus this work could be applied in heterogeneous application which employs materialized views.

Future scope of this research work includes replacement of some of the attributes (which are not in use) from the existing materialized views instead of replacing one materialized view at a time. However this require consideration of additional issues like joining of new attributes in views, discarding of attributes from views, considering the size constraints etc.